\documentclass{article}
\usepackage[utf8]{inputenc}
\usepackage{graphicx}
\usepackage{amsmath}
\usepackage{amsfonts}
\usepackage{amsthm}
\usepackage{hyperref}
\usepackage[ruled,linesnumbered,boxed,lined]{algorithm2e}
\usepackage{algpseudocode}
\usepackage{tikz}
\SetKwComment{Comment}{$\triangleright$ }{}
\SetKw{KwLet}{let}
\SetKw{KwYield}{yield}
\SetArgSty{}

\usepackage{lineno}

\theoremstyle{definition}
\newtheorem{definition}{Definition}
\newtheorem{theorem}{Theorem}[section]
\newtheorem{lemma}[theorem]{Lemma}

\title{Enumerating Unique Computational Graphs via an Iterative Graph Invariant}
\author{Chris Ying\\
        chrisying@google.com\\
        Google Brain}
        
\date{}

\begin{document}

\maketitle

\section{Introduction}

\begin{figure}
  \centering
  \begin{tikzpicture}[
      vertex/.style={
        fill opacity=0.25,
        text opacity=1,
        draw=black,
        circle,
        minimum width=0.5cm,
      },
      vertex 1/.style={vertex,fill=green},
      vertex 2/.style={vertex,fill=red},
      vertex 3/.style={vertex,fill=green},
      vertex 4/.style={vertex,fill=blue},
      vertex 5/.style={vertex,fill=blue},
      >=stealth,
    ]
    \begin{scope}
      \node [vertex 1] (A) at (+0, 0) {$1$};
      \node [vertex 2] (B) at (-1, 1) {$2$};
      \node [vertex 3] (C) at (+0, 2) {$3$};
      \node [vertex 4] (D) at (+1, 3) {$4$};
      \node [vertex 5] (E) at (+0, 4) {$5$};
      \draw [->]
        (A) edge (B)
        (A) edge (C)
        (A) edge (D)
        (B) edge (C)
        (C) edge (E)
        (D) edge (E)
        ;
    \end{scope}
    \begin{scope}[xshift=4cm]
      \node [vertex 1] (A) at (+0, 0) {$1$};
      \node [vertex 2] (B) at (-1, 1) {$2$};
      \node [vertex 3] (C) at (+1, 3) {$4$};
      \node [vertex 4] (D) at (+0, 2) {$3$};
      \node [vertex 5] (E) at (+0, 4) {$5$};
      \draw [->]
        (A) edge (B)
        (A) edge (C)
        (A) edge (D)
        (B) edge [bend left=40] (C)
        (C) edge (E)
        (D) edge (E)
        ;
    \end{scope}
    \begin{scope}[xshift=8cm]
      \node [vertex 1] (A) at (+0, 0) {$1$};
      \node [vertex 2] (B) at (+0, 2) {$3$};
      \node [vertex 3] (C) at (+1, 3) {$4$};
      \node [vertex 4] (D) at (-1, 1) {$2$};
      \node [vertex 5] (E) at (+0, 4) {$5$};
      \draw [->]
        (A) edge (B)
        (A) edge (C)
        (A) edge (D)
        (B) edge (C)
        (C) edge (E)
        (D) edge (E)
        ;
    \end{scope}
  \end{tikzpicture}
  \caption{%
    Three computational graphs with different adjacencies and colorings when ordered that are isomorphic in the sense of Definition~\ref{def:iso} (best viewed in color).
  }
  \label{fig:iso}
\end{figure}
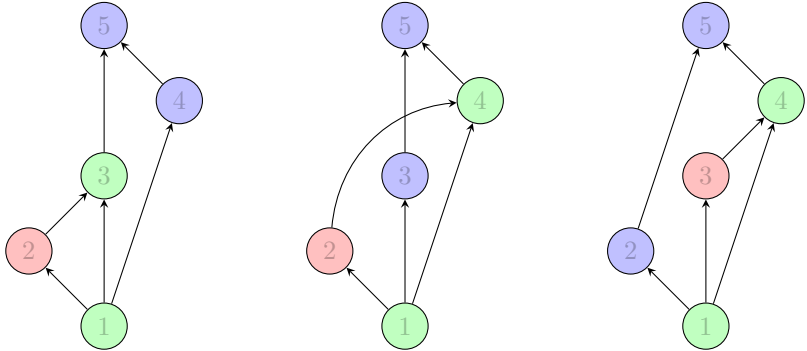

As part of a research project studying neural architectures, we needed an algorithm that could identify isomorphic \emph{computational graphs}, the building blocks of neural networks.
In our context, computational graphs represent operations performed on arbitrary tensors where the vertices are operations and the edges are tensors.
To generalize this problem, we represent a computational graph as a colored directed acyclic graph where the colors represent unique operations.
Depending on the graph representation (e.g., adjacency matrix plus a color per vertex), multiple representations may encode the same computational graph (Figure~\ref{fig:iso}).

In this report, we describe a novel graph invariant for computational graphs and how we used it to generate all distinct computational graphs up to isomorphism for small graphs. While this invariant cannot perfectly distinguish all pairs of non-isomorphic computational graphs, we suggest that it may be useful as a heuristic for comparing graphs.

\section{Definitions}

\begin{definition}
  A \emph{computational graph} on $n$~vertices and $k$~colors is a directed acyclic graph where each vertex is assigned an arbitrary color, and every vertex lies on a path between two designated vertices.
  Formally, it is a tuple $(n, k, E, c)$, where:
  \begin{enumerate}
    \item
      $n \in \mathbb{N}$ is the number of vertices;
    \item
      $k \in \mathbb{N}$ is the number of colors;
    \item
      $E \subseteq [n] \times [n]$ is a set of directed edges~$(i, j)$ such that~$i < j$;
    \item
      $c : [n] \to [k]$ is a function assigning a color to each vertex; and
    \item
      for each vertex~$i \in [n]$, there is a directed path from vertex~$1$ to vertex~$n$ that passes through vertex~$i$ \footnote{This condition is not required for the graph invariant algorithm~\ref{alg:igha} but is used by the enumeration algorithm~\ref{alg:enum} to heavily reduce the number of possible graphs.}.
  \end{enumerate}
\end{definition}
Note that there is no restriction that adjacent vertices have different colors.

The goal is to count the number of distinct computational graphs up to isomorphism, where isomorphism is defined as follows.
\begin{definition} \label{def:iso}
  Computational graphs $G_1 = (n, k, E_1, c_1)$ and $G_2 = (n, k, E_2, c_2)$ on $n$~vertices and $k$~colors are \emph{isomorphic} if there exists a bijection $P : [n] \to [n]$ such that:
  \begin{enumerate}
    \item
      adjacency is preserved: $\forall i, j \in [n].\ (i, j) \in E_1 \mathrel{\text{iff}} (P(i), P(j)) \in E_2$; and
    \item
      coloring is preserved: $\forall i \in [n].\ c_1(i) = c_2(P(i))$.
  \end{enumerate}
\end{definition}

\section{Related Approaches}

In the context of general graph isomorphism, a graph invariant is a property of a graph such that if two graphs are isomorphic, they have the same value of that property (the converse is not necessarily true). The Weisfeiler--Lehman algorithm \cite{weisfeiler_lehman_1968} uses an iterative coloring approach to come up with a canonical coloring on a graph, though subsequent work \cite{babai_kucera} shows that the algorithm can fail for some graphs. It is unknown whether graph isomorphism can be solved in polynomial time \cite{DBLP:journals/corr/Babai15}.

This paper deals with a more constrained problem than general graph isomorphism, where the graphs are directed acyclic graphs and there are colorings assigned to the vertices. Nonetheless, the algorithm is partially inspired by the iterative nature of the Weisfeiler--Lehman algorithm.

OEIS \cite{Sloane_theencyclopedia} provides a few sequences which are close to what we are looking for:
\begin{itemize}
    \item A000088: Number of graphs on $n$ unlabeled nodes. This series does not consider coloring and considers all graphs rather than just directed acyclic graphs.
    \item A003024: Number of acyclic digraphs with $n$ labeled nodes. This series treats all vertices as uniquely labeled rather than individual colored and also counts disconnected graphs.
    \item A057500: Number of connected labeled graphs with $n$ edges and $n$ nodes. This series does not use directed edges and does not consider coloring.
    \item A240955: Number of $k$-colored labeled digraphs with $n$ vertices. In this series, colored refers to vertices which cannot be colored the same color as its neighbors, which is different than the notion of coloring used here.
\end{itemize}

P\'{o}lya--Redfield counting can be used to count the number of colorings on undirected graphs \cite{robinson1968enumeration} but it does not provide a way to quickly identify if two directed colored graphs are isomorphic.

\section{Iterative Graph Hashing Algorithm}

In this section, we describe an algorithm for generating a novel graph invariant for computational graphs.
At a high-level, the key idea is to iteratively apply isomorphism-invariant operations to the graph in a way takes into account the graph structure as well as the coloring. 
Algorithm \ref{alg:igha} provides the pseudo-code for the graph hashing algorithm:
\\
\begin{algorithm}[H]
\KwIn{Computational graph $G = (n, k, E, c)$}
\KwOut{Fixed-length hash of the graph and coloring}
\SetAlgoLined
\KwLet $H = [\ ]$ \Comment*[r]{List such that $H_i$ is hash for vertex~$i$}
\ForAll{vertices~$i \in [n]$}{
    \KwLet $\lvert e^-_i \rvert = \texttt{in-degree}(i)$ \;
    \KwLet $\lvert e^+_i \rvert = \texttt{out-degree}(i)$ \;
    $H_i \gets \texttt{hash}(\lvert e^+_i \rvert, \lvert e^-_i \rvert, c(i))$ \Comment*[r]{Initialize hashes}
}
\For{$\text{iteration} \gets 1$ \KwTo $n$}{
    \KwLet $\hat{H} = [\ ]$ \Comment*[r]{Next iteration of hashes}
    \ForAll{vertices~$i \in [n]$}{
      \KwLet $H^-_i = [ H_o : (o, i) \in E ]$ \Comment*[r]{List of in-neighbor hashes}
      \KwLet $H^+_i = [ H_o : (i, o) \in E ]$ \Comment*[r]{List of out-neighbor hashes}
      $\hat{H}_i \gets \texttt{hash}(\texttt{sort}(H^+_i), \texttt{sort}(H^-_i), H_i)$ \;
    }
    $H \gets \smash{\hat{H}}$ \Comment*[r]{Update hashes}
}
\KwRet{$\texttt{hash}(\texttt{sort}(H))$}
\caption{Iterative graph hashing algorithm for colored DAGs}
\label{alg:igha}
\end{algorithm}
\vspace{1em}

Specifically, \texttt{hash} returns a fixed-length hash. Our implementation uses the 128-bit MD5 hash algorithm which we found sufficient for our use-case.
The \texttt{sort} function performs a lexicographical sort of the hash outputs.
We repeat the algorithm up to the number of vertices iterations (line 7) but we suspect that it may be sufficient to iterate up to the diameter of the graph.
In our implementation, we represent $G$ as an adjacency matrix along with a list of colors of length equal to the number of vertices.
The \texttt{in-degree} and \texttt{out-degree} functions are implemented as summations across the columns or rows of the matrix.

\subsection{Proof of graph invariance}

\begin{proof}[\unskip\nopunct]

To show that this algorithm computes a graph invariant, we must show that any two isomorphic any computational graphs output the same hash.
Consider two graphs $G = (n, k, E, c)$ and $G' = (n, k, E', c')$ that are isomorphic with isomorphism~$P$.
Suppose that we run Algorithm~\ref{alg:igha} on~$G$ and~$G'$.
We will use~$H$ and~$H'$ to refer to the values of the hashes in the executions of the algorithm on~$G$ and~$G'$, respectively.
We will say that the hashes are \emph{consistent} if they respect the isomorphism~$P$: that is, if~$H_i = H'_{P(i)}$ for each vertex~$i$.

Choose any vertex~$i$, and let~$j = P(i)$.
Then the initial hash~$H_i$ of~$i$ in~$G$ is equal to the initial hash~$H'_j$ of~$j$ in~$G'$ (after line~6), because the two vertices have the same adjacency and coloring, by Definition~\ref{def:iso}.
Thus:
\begin{lemma} \label{lem:1}
  The initial hashes~$H$ and~$H'$ are consistent (after line~6).
\end{lemma}

In each iteration of the outer for-loop (line~7), the hashes are updated.
Suppose, at the start of an iteration, that the hashes are consistent.
Again, choose any~$i$, and let~$j = P(i)$.
If $(o, i) \in E$, then $(P(o), P(i)) \in E'$ by the isomorphism condition.
By the consistency assumption, $H_o = H'_{P(o)}$.
Since $H^-_i = [ H_o : (o, i) \in E]$ and $H'^-_j = [H'_{P(o)} : (P(o), i) \in E']$ (line~10), then the \emph{multisets} (ignoring order) represented by $H^-_i$ and $H'^-_j$ are the same.
Likewise, $H^+_i$ is the same as $H'^+_j$ (line~11) ignoring order.
Because sorting ensures that the list orderings are identical, $\texttt{sort}(H^-_i) = \texttt{sort}(H'^-_j)$ and $\texttt{sort}(H^+_i) = \texttt{sort}(H'^+_j)$.
We also have $H_i = H'_j$ (because~$H$ and~$H'$ are consistent), so it follows that $\hat{H}_i = \hat{H'}_j$ (line~14) since the \texttt{hash} function is operating on the identical triplets.
Thus:
\begin{lemma} \label{lem:2}
  If the hashes are consistent at the start of an iteration (line~7), then they are also consistent at the end of that iteration (after line~14).
\end{lemma}

By induction, Lemma~\ref{lem:1} and Lemma~\ref{lem:2} show that the hashes are consistent throughout the full loop.
Because the final hashes~$H$ and~$H'$ are consistent, they are permutations of each other, so their sorted forms are the same, and thus the final hashed results are identical.
Therefore, Algorithm~\ref{alg:igha} computes a graph invariant.
\end{proof}

\section{Graph Enumeration Procedure}

Given the graph invariant, we can proceed to generate all computational graphs, up to isomorphism.
Using the canonical ordering, we treat the first vertex (no in-neighbors) as the ``input'' vertex and the last vertex (no out-neighbors) as the ``output'' vertex.

We observe that vertices not on a directed path from $v_1$ to $v_n$ in a colored directed acyclic graph can be pruned to yield a valid computational graph.
If we generate the directed acyclic graphs in increasing number of vertices order, then any graph that needs to be pruned has been already generated at a previous iteration and can be immediately skipped.
\\
\begin{algorithm}[H]
\newcommand*{\smax}{\mathrm{max}}
\KwIn{Maximum vertices~$n_\smax$, maximum edges~$e_\smax$, and colors~$k$}
\KwOut{Yields all unique computational graphs up to constraints}
\SetAlgoLined
\For{numbers of vertices~$n \gets 2$ \KwTo $n_{\smax}$}{
    \ForAll{bit vectors of length $n (n - 1) / 2$}{
        convert bit vector to $n \times n$ upper-triangular adjacency matrix \;
        \eIf{$\text{number of edges} > e_\smax$ \textbf{or} \\
        \quad contains vertex not on directed path from input to output}{
            discard and continue to next matrix \;
        }{
            \ForAll{potential colorings $c : [n] \to [k]$}{
                hash $(\text{matrix}, \text{coloring})$ using Algorithm~\ref{alg:igha} \;
                \If{hash has not been observed before}{
                    \KwYield $(\text{matrix}, \text{coloring})$ \;
                }
            }
        }
    }
}
\caption{Enumerating computational graphs}
\label{alg:enum}
\end{algorithm}
\vspace{1em}

This algorithm also provides a canonical computational graph (i.e., the first one that is observed) for each unique hash which represents the equivalence class of computational graphs induced by the Algorithm~\ref{alg:igha}.

\section{Verification}

For our neural network use-case, we needed to generate all graphs up to 7~vertices, 9~edges, and 3~colors.
Furthermore, the first vertex and the last vertex are specially colored and distinct from each other and the other 3~colors (they represent the input and output tensors of the network).

We verified that all graphs with the same hash generated in Algorithm~\ref{alg:enum} were unique up to isomorphism by running an expensive procedure which enumerates all possible permutations to confirm that any graph with a duplicate hash (line~11) is isomorphic to the canonical computational graph.
The definition of graph invariant implies that graphs with different hashes are non-isomorphic.

Thus for our constrained use-case, Algorithm~\ref{alg:igha} can exactly identify if two computational graphs are isomorphic or not.

\section{Adversarial graphs}

\begin{figure}
  \centering
  \begin{tikzpicture}[
      vertex/.style={
        fill opacity=0.25,
        text opacity=1,
        draw=black,
        circle,
        minimum width=0.5cm,
      },
      vertex 1/.style={vertex,fill=white},
      vertex 2/.style={vertex,fill=red},
      vertex 3/.style={vertex,fill=red},
      vertex 4/.style={vertex,fill=red},
      vertex 5/.style={vertex,fill=red},
      vertex 6/.style={vertex,fill=blue},
      vertex 7/.style={vertex,fill=blue},
      vertex 8/.style={vertex,fill=blue},
      vertex 9/.style={vertex,fill=blue},
      vertex 10/.style={vertex,fill=white},
      >=stealth,
    ]
    \begin{scope}
      \node [vertex 1] (1) at (1.5, 0) {$1$};
      \node [vertex 2] (2) at (0, 1) {$2$};
      \node [vertex 3] (3) at (1, 1) {$3$};
      \node [vertex 4] (4) at (2, 1) {$4$};
      \node [vertex 5] (5) at (3, 1) {$5$};
      \node [vertex 6] (6) at (0, 3) {$6$};
      \node [vertex 7] (7) at (1, 3) {$7$};
      \node [vertex 8] (8) at (2, 3) {$8$};
      \node [vertex 9] (9) at (3, 3) {$9$};
      \node [vertex 10] (10) at (1.5, 4) {$10$};
      \draw [->]
        (1) edge (2)
        (1) edge (3)
        (1) edge (4)
        (1) edge (5)
        (2) edge (6)
        (2) edge (7)
        (3) edge (7)
        (3) edge (8)
        (4) edge (8)
        (4) edge (9)
        (5) edge (9)
        (5) edge (6)
        (6) edge (10)
        (7) edge (10)
        (8) edge (10)
        (9) edge (10)
        ;
    \end{scope}
    \begin{scope}[xshift=6cm]
      \node [vertex 1] (1) at (1.5, 0) {$1$};
      \node [vertex 2] (2) at (0, 1) {$2$};
      \node [vertex 3] (3) at (1, 1) {$3$};
      \node [vertex 4] (4) at (2, 1) {$4$};
      \node [vertex 5] (5) at (3, 1) {$5$};
      \node [vertex 6] (6) at (0, 3) {$6$};
      \node [vertex 7] (7) at (1, 3) {$7$};
      \node [vertex 8] (8) at (2, 3) {$8$};
      \node [vertex 9] (9) at (3, 3) {$9$};
      \node [vertex 10] (10) at (1.5, 4) {$10$};
      \draw [->]
        (1) edge (2)
        (1) edge (3)
        (1) edge (4)
        (1) edge (5)
        (2) edge (6)
        (2) edge (7)
        (3) edge (6)
        (3) edge (7)
        (4) edge (8)
        (4) edge (9)
        (5) edge (8)
        (5) edge (9)
        (6) edge (10)
        (7) edge (10)
        (8) edge (10)
        (9) edge (10)
        ;
    \end{scope}
  \end{tikzpicture}
  \caption{%
    A counterexample using 10 vertices and 16 edges. Vertices 2, 3, 4, 5 must be the same color and likewise for 6, 7, 8, 9 (the two sets of vertices can be the same color). Vertices 1 and 10 can be colored with any color. 
  }
  \label{fig:counter}
\end{figure}
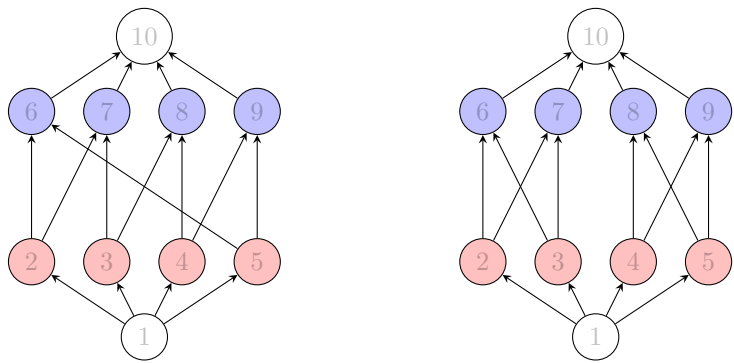

An adversarial example to the identifiability of the algorithm consists of two non-isomorphic computational graphs which hash to the same value. One such example can be seen in Figure~\ref{fig:counter} using 10 vertices and 16 edges. The counterexample holds also long as vertices 2, 3, 4, 5 are the same color and likewise for vertices 6, 7, 8, 9. The two graphs are non-isomorphic by inspection but Algorithm~\ref{alg:igha} fails to distinguish between the two. This is because vertices 2, 3, 4, 5 all start with the same initial hash due to having the same degree and each iteration maintains this equivalence because the in and out neighbors share the same colors (and likewise for vertices 6, 7, 8, 9).

An infinite number of similar adversarial graphs can be constructed from pairs of directed non-isomorphic bipartite graphs where all edges point from one partition to the other and the degree of all vertices within each partition is the same.

\section{Future Work}

Modifying the algorithm to deal with cases like the counterexample above is the first direction for future work.

In addition to the counter example discussed above, another possible problem is hash collision. A possible solution is to replace the \texttt{hash} function with string concatenation, which would cause the iterative ``hashes'' to grow exponentially in length at each iteration.
This eliminates the possibility of hash collision, and the proof of graph invariance still holds.

\subsection*{Acknowledgements}

We would like to thank Chris Jones for suggesting Weisfeiler--Lehman color refinement and finding a counterexample, Esteban Real for reviewing the code implementation of the algorithm, and William Chargin for reviewing the notation and proof.

\bibliography{bibliography}
\bibliographystyle{plain}

\end{document}